# Distributed UAV Swarm Augmented Wideband Spectrum Sensing Using Nyquist Folding Receiver

Kaili Jiang, Kailun Tian, Hancong Feng, Yuxin Zhao, Dechang Wang
Sen Cao, Jian Gao, Xuying Zhang, Yanfei Li, Junyu Yuan, Ying Xiong and Bin Tang

*Abstract*—Distributed unmanned aerial vehicle (UAV) swarms are formed by multiple UAVs with increased portability, higher levels of sensing capabilities, and more powerful autonomy. These features make them attractive for many recent applications, potentially increasing the shortage of spectrum resources. In this paper, wideband spectrum sensing augmented technology is discussed for distributed UAV swarms to improve the utilization of spectrum. However, the sub-Nyquist sampling applied in existing schemes has high hardware complexity, power consumption, and low recovery efficiency for non-strictly sparse conditions. Thus, the Nyquist folding receiver (NYFR) is considered for the distributed UAV swarms, which can theoretically achieve full-band spectrum detection and reception using a single analog-to-digital converter (ADC) at low speed for all circuit components. There is a focus on the sensing model of two multichannel scenarios for the distributed UAV swarms, one with a complete functional receiver for the UAV swarm with RIS, and another with a decentralized UAV swarm equipped with a complete functional receiver for each UAV element. The key issue is to consider whether the application of RIS technology will bring advantages to spectrum sensing and the data fusion problem of decentralized UAV swarms based on the NYFR architecture. Therefore, the property for multiple pulse reconstruction is analyzed through the Gershgorin circle theorem, especially for very short pulses. Further, the block sparse recovery property is analyzed for wide bandwidth signals. The proposed technology can improve the processing capability for multiple signals and wide bandwidth signals while reducing interference from folded noise and subsampled harmonics. Experiment results show augmented spectrum sensing efficiency under non-strictly sparse conditions.

Manuscript received July 3, 2023; revised month date, year; accepted month date, year. Date of publication month date, year; date of current version month date, year. This work was supported in part by the Key Project of the National Defense Science and Technology Foundation Strengthening Plan 173 under Grand 2022-JCJQ-ZD-010-12. The associate editor coordinating the review of this manuscript and approving it for publication was xxx. (*Corresponding author: Kaili Jiang*).

Kaili Jiang, Yuwei Huang, Kailun Tian, Hancong Feng, Yuxin Zhao, Junyu Yuan, Ying Xiong and Bin Tang are with the School of Information and Communication Engineering, University of Electronic Science and Technology of China, Chengdu, Sichuan, 611731, China (e-mail: jiangkelly@uestc.edu.cn, kailun_tian@163.com, 2927282941@qq.com, 1051172535@qq.com, c13844033835@163.com, Jyyuan@uestc.edu.cn, yxiong@uestc.edu.cn , bint@uestc.edu.cn).

Sen Cao, Jian Gao, Xuying Zhang and Yanfei Li are with the China Electronics Technology Group Corporation 29th Research Institute, Chengdu, Sichuan, 610036, China (e-mail: caosen@cetc. com.cn, swieegaoj@sina. com, zhxuying @126.com, lyf286lyf @sina.com).

## I. INTRODUCTION

INTERNET of Things (IoT) has rapidly grown in the past few years, and UAVs form an important part of it. With their unique features, UAVs have great application potential in the military and civilian domains, such as emergency response, remote sensing, transportation, agricultural spraying, surveillance and even attacking. As their size and weight are further reduced and their sensing capabilities and level of autonomy are further enhanced, which make UAVs a perfect choice for applications in extreme conditions [1]. In addition, UAV swarms comprised of several UAVs are attracting more interest, which can collaboratively perform tasks more robustly, economically, and efficiently than a single one [2].

Recently, two mainstream schemes of distributed UAV swarms have been widely considered, which are reconfigurable intelligent surfaces (RIS) assisted UAV swarm and the decentralized UAV swarm. The UAV swarm with RIS is a low-cost system that uses RIS elements with only one complete functional receiver for the swarm [3]. The RIS has been widely applied due to the channel reconstruction ability and cost-effectiveness [4-5]. Meanwhile, the capabilities have been enhanced by reflecting received signals with controllable reflection amplitudes and phases for applications such as DOA estimation [6-7] and wireless communication [8-9]. Moreover, compared to this centralized scheme, the decentralized UAV swarm is equipped with a complete functional receiver for each UAV element [10-11], enabling the UAV swarm to complete the tasks with a high probability even when some UAV elements are partially damaged. Thus, both scenarios have their advantages that will complement each other in future development.

With the rapid application of UAV technology, the number of IoT devices is expected to grow substantially, making it difficult to allocate enough spectrum bands for these devices [12]. Such an increase magnifies the scarcity of radio frequency (RF) spectrum resources, becoming a more critical problem for the future of IoT [13]. Therefore, a fundamental technique is needed to identify the frequency locations of multiple signals over a wide frequency band, which can be applied to electronic surveillance [14], cognitive radio [15], and broadband power line communication (BPLC) [16], among others. Furthermore, wideband spectrum sensing has been and will continue to be important in the signal processing domain for a long time.

As the monitoring frequency band range increases, the difficulty of acquiring signal parameters also increases. On the one hand, the sampling rate and dynamic range of the existing ADC are difficult to meet the application requirements based on the Nyquist-Shannon sampling theorem [17]. On the other hand, there is a great pressure on processing, transmitting, and storing massive amounts of data. To overcome these difficulties, the bandpass sampling and the sub-Nyquist sampling have been presented over the past decade.

The conventional wideband reception architecture scans the frequency spectrum sequentially based on the number of separate frequency segments being monitored. Nevertheless, such a scanning scheme has a long scan time and a low intercept probability for short-lived pulses [18]. And the most practical wideband reception architecture currently is the channelized scheme. However, there are several application problems [19] such as a large number of ADCs, high power consumption, complex hardware structure, and serious RF crosstalk. Both of these schemes are possible to sample the inputs with carrier frequency greater than the sampling rate, although the bandwidth occupied by the received signals is limited by the sampling theorem to less than half of the sampling rate.

To overcome the limitations imposed by the sampling theorem, there have been studies to achieve high-fidelity reconstruction of subsampled signals. These methods often involve additional constraints, such as sparsity in a certain domain or regular nonuniformity in the sampling process. However, these techniques come at the cost of increased algorithmic complexity and computational effort.

The addition of non-uniform constraints to the signal sampling process can be achieved through multi-channel time domain splicing sampling, which includes multi-rate sampling (MRS) and multi-coset sampling (MCS) frameworks. A representative structure of the MRS framework is the coprime sampling structure [20-21]. However, extending the number of detectable signals comes at the cost of increased hardware consumption. And the representative structures of the MCS framework include the time interleaved ADC (TIADC) structure and the periodic non-uniform sampling structure [22-23]. The TIADC structure is currently a preferred solution for high-speed, high-precision sampling systems as it effectively increases the sampling rate. However, it fails to achieve a high effective bit due to the systematic error caused by channel mismatching [24]. The calibration problem of time delay difference can be equated to the reconstruction problem of periodically non-uniformly sampled signals [25].

The non-uniform sampling structures have the advantage of easy hardware implementation mentioned above, but still does not break through the theoretical framework of Nyquist sampling. In contrast, the recent direct application of compressive sensing (CS) [26-28] or deep compressive sensing (DCS) [29-31] theory to the signal sampling process represents a fundamental breakthrough of the sampling theorem, which requires the structure or content of the sampled signal to satisfy the sparsity constraint.

The physical implementation schemes for CS-based signal acquisition are collectively referred to as analog information conversion techniques [32]. The representative structures include random sampling (RS), random demodulation (RD), and the modulated wideband converter (MWC). The RS structure uses the randomness of the sampling clock to preserve the spectral information in each sample to varying degrees. The RD structure achieves low-rate acquisition of wideband signals by multiplying them with a random Nyquist rate sequence in the time domain as a linear transformation. However, the randomness of the sampling clock and sequence is difficult to be guaranteed in practice. Until now, the MWC structure has received most attention, which mixes signals with multiple cycles of pseudo-random Nyquist rate sequences and mixes all signals into the baseband with the low-pass filtering to achieve analog compression of wideband signals. However, designing the mixing function for each channel is very challenging [33].

In particular, the NYFR structure is a low-complexity architecture within the CS framework that can theoretically achieve full-band signal detection and reception using a single-channel technique [34-37]. By mapping the Nyquist Zone (NZ) information of inputs to a modulation bandwidth parameter, the wideband reception of multi-signals is achieved using a single low-speed ADC device. Moreover, only the NZ index value of the signal needs to be estimated, after which the existing signal processing methods can be used to obtain the subsampled signal information. As a result, the NYFR structure is more suitable for UAV platforms than others, which has higher engineering feasibility.

Existing studies on the NYFR structure have primarily focused on the continuous wave (CW) signals [34-36]. In this paper, pulse signals are the focus, particularly for the short pulse that can be recovered as the boundary conditions. Additionally, the existing studies have not considered the case of multiple reception channels. The single channel structure has issues such as interference with folded noise and harmonics, reduced processing capability for the multiple signals and cross-Nyquist zone (NZ) signals, and others based on the inherent bandwidth broadening characteristics of the received signal [37]. There will be a focus on the sensing model of two multichannel scenarios for the distributed UAV swarms and analyze the restricted isometry property (RIP) for multiple signal reconstruction. In particular, for wide bandwidth signals and cross-NZ signals, the block-restricted isometry property (BRIP) is analyzed deeply for block sparse recovery. Finally, the influence of the aliasing on the subsampled inputs, and the degree of model matching for the system are discussed.

The rest of this paper is organized as follows. Section II describes the system scenario and signal model. In section III, the sensing model is formulated for the two scenarios, and the RIP and BRIP of the sensing model are discussed based on the Gershgorin circle theorem. Section IV analyzes and validates the proposed scheme through simulation. The conclusion is presented in section V.

## II. PROPOSED SYSTEM SCENARIO

In this paper, two typical scenarios are considered in the target spectrum ranges of 2-18GHz. One is that the UAV swarm has a center UAV and multiple UAVs with RIS element. The central UAV receives signals reflected from the UAV swarm with RIS using the NYFR. The other is a distributed UAV swarm configured with the NYFR for each.

*A. UAV Swarm with RIS using NYFR*

Consider a wideband spectrum sensing scenario that uses $M$ UAVs, each equipped with an RIS element. The UAVs are arranged in a uniform linear array (ULA), and a central UAV receives the signals reflected from the RIS elements. The schematic of the UAV swarm with RIS using NYFR is shown in Fig. 1. The UAV swarm has a fully functional receiving system, mounted on the central UAV, while the remaining UAVs are equipped with the low-cost RIS elements.

Assume that the position of the $m$th UAV is denoted as $d_m$, with the position of the 0th UAV serving as the reference position, that is $d_0 = 0$ for $m=0,1,\ldots,M-1$. For the $m$th UAV, the $K$ uncorrelated and far-field received signals over a wide frequency band of wavelength $\lambda_k$ colliding into the UAV swarm can be expressed as

$$r_m(t) = \sum_{k=0}^{K-1} s_k(t) e^{j2\pi \frac{d_m}{\lambda_k}\sin\varphi_k} \triangleq \sum_{k=0}^{K-1} r_{m,k}(t) \quad (1)$$

where $s_k(t)$ and $\varphi_k$ denote the signal of interest and the direction of the $k$th signal with $k = 0,1,\ldots,K-1$, respectively. The signals have different carrier frequencies $f_{c_k}$ that enable them to be distinguished from each other, with these frequencies confined within the bandwidth $B_s$. And the unknown directions $\varphi_k$ are limited to a range of $(-\pi/3, \pi/3]$, which are represented as only a phase in the UAVs based on the far-field model of signal propagation. After reflection by the RIS, the signal received by the central UAV is as follows

$$\begin{aligned} x(t) &= \sum_{m=0}^{M-1}\sum_{k=0}^{K-1} A_m e^{j\alpha_m} r_{m,k}(t) e^{j2\pi \frac{d_m}{\lambda_k}\sin\beta} + w(t) \\ &\triangleq \sum_{m=0}^{M-1}\sum_{k=0}^{K-1} x_{m,k}(t) + w(t) \end{aligned} \quad (2)$$

where $A_m$ and $\alpha_m$ represent the reflection amplitude and phase caused by the $m$th RIS, respectively. While the $\beta$ denotes the know far-field direction between the central UAV and the UAV swarm. Finally, the $w(t)$ denotes the additive white Gaussian noise (AWGN) with the variance $\sigma_w^2$.

We attempt to reconstruct the spectrum of signals received from the low-cost system, which are sampled using the NYFR. The diagram of the NYFR architecture is shown in Fig. 2. The inputs are obtained through direct RF pulse-based sampling without quantization, after passing through the preselected band pass filter (BPF) and a low noise amplifier (LNA) for the entire frequency range.

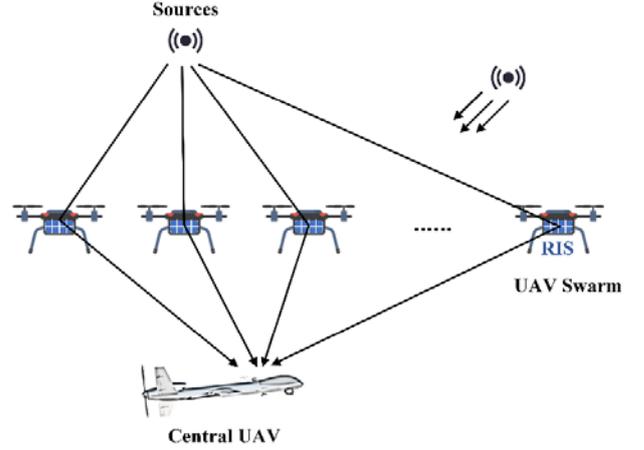

**Fig. 1.** Scenario of wideband spectrum sensing using a UAV swarm with RIS.

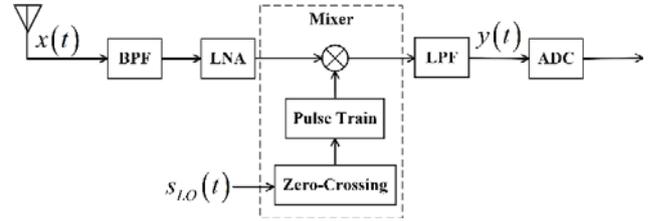

**Fig. 2.** Diagram of the Nyquist folding receiver architecture.

The RF pulses are created at each positive zero crossing of the reference local oscillator (LO) signal $s_{LO}(t)$ as follows

$$s_{LO}(t) = \sin(2\pi f_s t + \theta(t)) \quad (3)$$

which is a frequency modulated signal centered at $f_s$ with the phase modulation $\theta(t)$ and that is a sinusoidal modulation defined as

$$\theta(t) = A_\theta \sin(2\pi f_\theta t) \quad (4)$$

where $A_\theta$ and $f_\theta$ is the amplitude and frequency of the sinusoidal modulation signal, respectively. Meanwhile, the RF pulse-based sampling is periodic and nonuniform, with a period of $T = 1/f_s$. The output of the anti-aliasing low pass filter (LPF) after the harmonic mixing is as follows

$$y(t) = \sum_{m=0}^{M-1}\sum_{k=0}^{K-1} x_{m,k}(t) e^{jM_k(2\pi f_s t + A_\theta \sin(2\pi f_\theta t))} + w(t) \quad (5)$$

where $M_k$ is an integer for the induced Nyquist zone index and that satisfies $0 \leq |f_{c_k} - M_k f_s| \leq 0.5 f_s$ with the cutoff frequency of $0.5 f_s$ for the LPF. The output is digitized by the conventional ADC at last.

Finally, from the system model (5), we try to recovery the interested spectrum information of $f_{c_k}$ form the received signal $y(t)$ in the scenario for the UAV swarm with RIS using NYFR. Moreover, the wideband spectrum sensing of multiple targets, the multiple measurements are realized by the sensing matrix which is further introduced in the next section.

*B. Decentralized UAV Swarm using NYFR*

Consider a wideband spectrum sensing scenario that uses $P$ UAVs, where the UAV swarm can be arranged in a non-uniform array. The schematic of the distributed UAV swarm using NYFR is shown in Fig. 3. It is assumed that $K$ uncorrelated and far-field signals collide with the UAV swarm over a wide frequency band. Moreover, the process is wide-sense stationary with zero mean. Let $x_p(t), t \in \mathbb{R}$ combine the $K$ signals from the $p$ th UAV in the time domain, which can be expressed as

$$x_p(t) = \sum_{k=0}^{K-1} s_k(t - \tau_{k,p}) + w_p(t) \quad (6)$$

where $s_k(t)$ and $\tau_{k,p}$ denote the RF signal of interest and its time delay in the $p$ th UAV of the $k$ th source, respectively. And the term $w_p(t)$ denotes the additive white Gaussian noise.

In the far-field propagation model, the time delay in the UAV swarm is considered. Meanwhile, each source signal has an unknown azimuth $\theta_i \in [-\pi/3, \pi/3]$ corresponding to the time delay. And the signals have distinct carrier frequencies $f_{c_i}$ confined within the bandwidth $B_s$ at one instant, which allows them to be distinguished from one another. Furthermore, the AWGN is an independent and identically distributed random variable, with zero mean and variance $\sigma_w^2$.

The output samples of the UAV swarm are obtained by $P$ NYFRs with different RF pulse-based sampling, and there is the same periodic sampling rate. It means that the receivers have the same system parameters except the phase modulation for the LO signals. And the diagram for the distributed NYFR architecture is shown in Fig. 4. Thus, the received outputs with the $P$ samplers can be respectively written as

$$y_p(t) = \sum_{k=0}^{K-1} s_k(t - \tau_{k,p}) e^{jM_k\left(2\pi f_s t + A_{\theta,p} \sin(2\pi f_\theta,p t)\right)} + w_p(t) \quad (7)$$

Compared to the schematic of a UAV swarm with RIS using NYFR, the distributed UAV swarm using NYFR is a problem of the multichannel signal reconstruction. The difference is the entire spectrum sampled by each UAV compared to the earlier studies. It means that the received target signals are the same, but the received outputs are different within the same observation interval. Therefore, the focus of this work is to construct sensing matrix for the multichannel architecture and reconstruct the interested spectrum information.

## III. WIDEBAND SPECTRUM RECOVERY

In this section, the CS model is first constructed for the two scenarios described above. The UAV swarm with RIS using NYFR is a typical example of the problem of single-channel signal reconstruction. While the distributed UAV swarm using NYFR is a typical example of the problem of multichannel signal reconstruction. The RIP and BRIP are then analyzed in detail in the following.

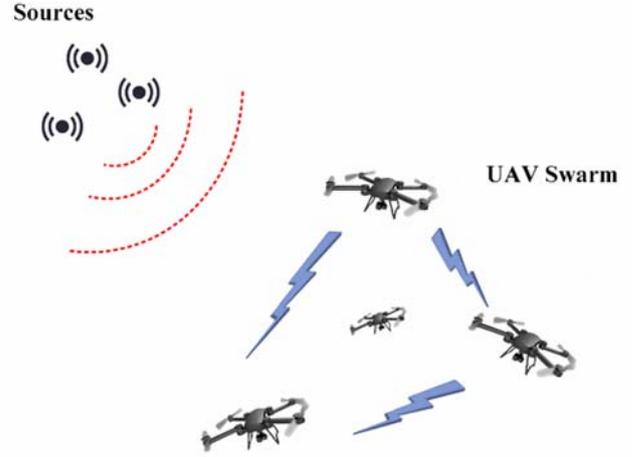

**Fig. 3.** Scenario of wideband spectrum sensing using a distributed UAV swarm.

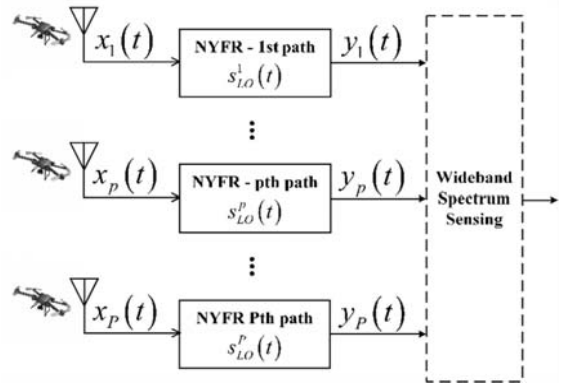

**Fig. 4.** Diagram for the distributed Nyquist folding receiver architecture.

*A. Compressive Sensing Model*

A sensing matrix of the multiple measurement vectors is first designed for the UAV swarm with RIS using NYFR. By collecting all $N$ measurements, the measurement signal can be expressed as

$$\mathbf{y} \triangleq \left[y(0), y(T), \ldots, y((N-1)T)\right]^{\mathrm{T}} \in \mathbb{C}^{N \times 1} \quad (8)$$

where

$$y(nT) = \mathbf{B}^{\mathrm{T}} \mathbf{A}(\psi, \varphi, \mathbf{d}) \Theta(nT) \mathbf{s}(nT) + w(nT) \quad (9)$$

for $n = 0, 1, \ldots, N-1$. The signal vector is defined as

$$\mathbf{s}(nT) = \left[s_0(nT), s_1(nT), \ldots, s_{K-1}(nT)\right]^{\mathrm{T}} \in \mathbb{C}^{K \times 1} \quad (10)$$

And the modulation matrix $\Theta(nT)$ is defined as

$$\Theta(nT) = diag \begin{pmatrix} e^{jM_0\left(2\pi f_s(nT) + A_\theta \sin(2\pi f_\theta(nT))\right)} \\ e^{jM_1\left(2\pi f_s(nT) + A_\theta \sin(2\pi f_\theta(nT))\right)} \\ \vdots \\ e^{jM_{K-1}\left(2\pi f_s(nT) + A_\theta \sin(2\pi f_\theta(nT))\right)} \end{pmatrix} \in \mathbb{C}^{K \times K} \quad (11)$$

which is based on the NYFR architecture with the reference LO signal $s_{LO}(t)$ and the phase modulation $\theta(t)$.

The steering matrix $\mathbf{A}(\psi,\boldsymbol{\varphi},\mathbf{d}) \in \mathbb{C}^{M \times K}$ is then defined with the direction $\psi$, the vector $\boldsymbol{\varphi} \triangleq [\varphi_0, \varphi_1, \ldots, \varphi_{K-1}]^T \in \mathbb{C}^{M \times K}$ and the position $\mathbf{d}$ as

$$\mathbf{A}(\psi,\boldsymbol{\varphi},\mathbf{d}) \triangleq [\mathbf{a}(\psi,\varphi_0,\mathbf{d}), \mathbf{a}(\psi,\varphi_1,\mathbf{d}), \ldots, \mathbf{a}(\psi,\varphi_{K-1},\mathbf{d})] \quad (12)$$

where the steering vector is defined as

$$\mathbf{a}(\psi,\varphi_k,\mathbf{d}) = \begin{bmatrix} e^{j2\pi \frac{d_0}{\lambda_k}(\sin\varphi_k + \sin\psi)} \\ e^{j2\pi \frac{d_1}{\lambda_k}(\sin\varphi_k + \sin\psi)} \\ \vdots \\ e^{j2\pi \frac{d_{M-1}}{\lambda_k}(\sin\varphi_k + \sin\psi)} \end{bmatrix} \in \mathbb{C}^{M \times 1} \quad (13)$$

At last, the measurement matrix $\mathbf{B}$ is defined as

$$\mathbf{B} = \left[ A_0 e^{j\alpha_0}, A_1 e^{j\alpha_1}, \ldots, A_{M-1} e^{j\alpha_{M-1}} \right] \in \mathbb{C}^{M \times 1} \quad (14)$$

Therefore, according to the signal model in (5), the sensing model for spectrum recovery based on the CS is defined as

$$\mathbf{y} = \mathbf{H}\mathbf{X} + \mathbf{W} \quad (15)$$

where $\mathbf{W} = [w(0), w(T), \ldots, w((N-1)T)]^T$ is the AWGN vector of the NYFR output and $\mathbf{X} = [X_0, X_1, \ldots, X_{ZN}]^T$ is the discrete Fourier transform (DFT) of the Nyquist-rate sampled signal $x(t)$, which is sparse or compressible in the frequency domain.

Due to the linearity property of the measurement matrix $\mathbf{B}$ and time shifting for the steering matrix $\mathbf{A}(\psi,\boldsymbol{\varphi},\mathbf{d})$ of the received signal in the central UAV, its DFT $\mathbf{X} \in \mathbb{C}^{ZN}$ shares the same spectrum feature with the interested signal $s_k(t)$ for $k = 0, 1, \ldots, K-1$. Moreover, the spectrum of $\mathbf{X}$ consists of $Z$ Nyquist zones with each of length $N$. Then the sensing matrix $\mathbf{H} \in \mathbb{C}^{N \times ZN}$ is

$$\mathbf{H} = \mathbf{P}\boldsymbol{\Theta}\boldsymbol{\Psi} \quad (16)$$

where $\boldsymbol{\Psi} \in \mathbb{C}^{ZN \times ZN}$ is the block diagonal matrix of the inverse discrete Fourier transform (IDFT), where each block denotes a Nyquist zone, such that

$$\boldsymbol{\Psi} = \mathbf{I}_Z \otimes \boldsymbol{\Psi}_N \quad (17)$$

where $\mathbf{I}_Z \in \mathbb{C}^{Z \times Z}$ has the same dimension as the number of Nyquist zones. And $\boldsymbol{\Psi}_N \in \mathbb{C}^{N \times N}$ is the IDFT matrix with the rotation factor $\psi = e^{j2\pi/N}$. Then the induced sampling modulation matrix $\boldsymbol{\Theta} \in \mathbb{C}^{ZN \times ZN}$ is a diagonal matrix with $Z$ blocks, that is

$$\boldsymbol{\Theta} = diag(\boldsymbol{\Theta}_0, \boldsymbol{\Theta}_1, \ldots, \boldsymbol{\Theta}_{Z-1}) \quad (18)$$

where $\boldsymbol{\Theta}_z \in \mathbb{C}^{N \times N}$ is the discrete sampled signal of the phase modulation $\theta(t)$ at the ADC sampling rate, which is also a diagonal matrix

$$\boldsymbol{\Theta}_z = diag\left(1, e^{jM_z A_\theta \sin(2\pi f_\theta T)}, \ldots, e^{jM_z A_\theta \sin(2\pi f_\theta (N-1)T)}\right) \quad (19)$$

where $M_z = round(f_z/f_s)$ is the induced modulation index for each Nyquist zone frequency $f_z$ with $z = 0, 1, \ldots, Z-1$.

Finally, the projection matrix $\mathbf{P} \in \mathbb{C}^{N \times ZN}$ folds the $Z$ Nyquist zones into one zone by horizontally concatenating the identity matrices, such that

$$\mathbf{P} = \mathbf{e}_Z \otimes \mathbf{I}_N \quad (20)$$

where $\mathbf{e}_Z \in \mathbb{C}^{1 \times Z}$ is a vector with all one elements.

The sensing matrix with the multichannel measurement vectors is then designed for the distributed UAV swarm using NYFR. By stacking all received vectors $\mathbf{y}_p \in \mathbb{C}^{N \times 1}$ into a vector $\mathbf{y}''$ of size $PN \times 1$ vertically, the sensing model based on the CS according to the signal model in (7) for spectrum recovery is defined as

$$\underbrace{\begin{pmatrix} \mathbf{y}_1 \\ \mathbf{y}_2 \\ \vdots \\ \mathbf{y}_P \end{pmatrix}}_{\mathbf{y}'} = \underbrace{\begin{pmatrix} \mathbf{P} & & & \\ & \mathbf{P} & & \\ & & \ddots & \\ & & & \mathbf{P} \end{pmatrix}}_{\mathbf{P}'} \underbrace{\begin{pmatrix} \boldsymbol{\Theta}^1 \\ \boldsymbol{\Theta}^2 \\ \vdots \\ \boldsymbol{\Theta}^P \end{pmatrix}}_{\boldsymbol{\Theta}'} \boldsymbol{\Psi}\mathbf{X} + \mathbf{W} \quad (21)$$

where $\mathbf{W}$ and $\mathbf{X}$ are the AWGN vector and the DFT vector of the Nyquist-rate sampled signal $x(t)$ with the same as (15), respectively. The block diagonal matrix $\boldsymbol{\Psi}$ of IDFT is the same as (17) of size $ZN \times ZN$. Then the induced sampling modulation matrix $\boldsymbol{\Theta}' \in \mathbb{C}^{PZN \times ZN}$ is obtained by stacking the $\boldsymbol{\Theta}^P \in \mathbb{C}^{ZN \times ZN}$ into a matrix vertically, which has the same structure as (18) but with a different modulation pattern for each one. The remainder of the projection matrix $\mathbf{P}' \in \mathbb{C}^{PN \times PZN}$ becomes a block diagonal matrix, where each block is the same as (20) and each one is for each channel.

The relationship between the spectrum of the UAV swarm with RIS or the distributed UAV swarm and the NYFR output follows the sensing model (15) and (21), respectively. Then, the spectrum of interest can be reconstructed by solving the convex optimization problem as follows

$$\begin{aligned} & \text{minimize} & & \|\mathbf{X}\|_0 \\ & \text{subject to} & & \mathbf{y} = \mathbf{H}\mathbf{X} + \mathbf{W} \end{aligned} \quad (22)$$

or

$$\begin{aligned} & \text{minimize} & & \|\mathbf{X}\|_0 \\ & \text{subject to} & & \mathbf{y}' = \mathbf{P}'\boldsymbol{\Theta}'\boldsymbol{\Psi}\mathbf{X} + \mathbf{W} \end{aligned} \quad (23)$$

where the minimization problem of $l_0$ norm for a vector is a NP-hard problem, and cannot be solved directly. Fortunately, the problem can be transformed from a non-convex problem to a convex problem, and then the approximation can be found to satisfy the accuracy requirement.

The received signal of the time domain $\mathbf{x} \in \mathbb{C}^{ZN \times 1}$ based on the Nyquist-rate sampling can be reconstructed through

$$\mathbf{x} = \boldsymbol{\Psi}_{ZN}\mathbf{X}_{\text{reco}} \quad (24)$$

where the $\boldsymbol{\Psi}_{ZN} \in \mathbb{C}^{ZN \times ZN}$ is the IDFT matrix with the rotation factor $\psi = e^{j2\pi/ZN}$. And $\mathbf{X}_{\text{reco}} \in \mathbb{C}^{ZN \times 1}$ is the reconstructed spectrum from the sensing model.

The typical reconstruction algorithms include convex optimization, greedy iteration, Bayesian learning, etc. However, the high probability of recovery has to be satisfied by the RIP of the sensing matrix, which is analyzed in the following.

*B. The RIP analysis*

The RIP is defined as follows: a matrix $\mathbf{H} \in \mathbb{C}^{N \times ZN}$ is said to satisfy the RIP with parameters $(s, \delta)$ for $s \leq N, 0 \leq \delta \leq 1$, if for all subsequences of the index set $I \subset \{1, 2, \ldots, ZN\}$ of $\mathbf{H}$ such that $|I| \leq s$, and for all $\boldsymbol{\theta} \in \mathbb{C}^{|I|}$, one has

$$(1-\delta)\|\boldsymbol{\theta}\|_2^2 \leq \|\mathbf{H}_I \boldsymbol{\theta}\|_2^2 \leq (1+\delta)\|\boldsymbol{\theta}\|_2^2 \quad (25)$$

where $\mathbf{H}_I$ is the submatrix of $\mathbf{H}$ consisting of the related columns specified by the index set $I$ of size $N \times |I|$. Then, the parameter $s$ denotes the sparsity and the minimum of all $\delta$ is the restricted isometry constant (RIC) $\delta_s$. There is a relation of inequality between the RIC and the eigenvalues of the matrix $\mathbf{H}_I^H \mathbf{H}_I$, as follows

$$1 - \delta_s \leq \lambda_{\min}\left(\mathbf{H}_I^H \mathbf{H}_I\right) \leq \lambda_{\max}\left(\mathbf{H}_I^H \mathbf{H}_I\right) \leq 1 + \delta_s \quad (26)$$

where $\lambda_{\min}\left(\mathbf{H}_I^H \mathbf{H}_I\right)$ and $\lambda_{\max}\left(\mathbf{H}_I^H \mathbf{H}_I\right)$ denote the minimal and maximal eigenvalues of $\mathbf{H}_I^H \mathbf{H}_I$, respectively.

Then, the Gershgorin circle theorem is used to understand the RIP. Considering the sensing matrix $\mathbf{H} = \mathbf{P\Theta\Psi}$ as (16), there are $C_{ZN}^s$ submatrices $\mathbf{H}_I$ by randomly selecting columns of size $s$. Thus, the eigenvalues of each submatrix of the Gram matrix $\mathbf{H}_I^H \mathbf{H}_I$ are distributed in $[1-\delta_s, 1+\delta_s]$, which is a complicated permutation and combination problem. According to the Gershgorin circle theorem, the Gram matrix of $\mathbf{H}$ contains all the eigenvalue information of its submatrices. Therefore, it is reasonable to analyze the Gram matrix $G(\mathbf{H})$, which is given by

$$G(\mathbf{H}) = \mathbf{H}^H \mathbf{H} = \mathbf{\Psi}^H \left(\mathbf{\Theta}^H \left(\mathbf{P}^H \mathbf{P}\right) \mathbf{\Theta}\right) \mathbf{\Psi}$$
$$= \begin{bmatrix} \mathbf{I}_N & \mathbf{T}_{10} & \cdots & \mathbf{T}_{(Z-1)0} \\ \mathbf{T}_{01} & \mathbf{I}_N & \cdots & \mathbf{T}_{(Z-1)1} \\ \vdots & \vdots & \ddots & \vdots \\ \mathbf{T}_{0(Z-1)} & \mathbf{T}_{1(Z-1)} & \cdots & \mathbf{I}_N \end{bmatrix} \quad (27)$$

where the square matrix $\mathbf{T}_{ij}$ $(i, j = 0, 1, \ldots, Z-1; i \neq j)$ of size $Z \times Z$ is expressed as (30), where $M_{ij} = M_i - M_j$, the $M_i$ means $i$ th Nyquist zone and $M_j$ is the same.

To simplify the matrix (30), we have

$$\mathbf{T}_{ij} = \left(\frac{1}{N} \sum_{n=1}^{N} e^{-jM_{ij}A_\theta\theta(nT)} \psi^{(n-1)(r-c)}\right)_{r,c} \quad (28)$$

where $r$ and $c$ denote the row and column of the matrix $\mathbf{T}_{ij}$ within $r, c = 1, 2, \ldots, N$, respectively.

After considering the multichannel sensing matrix $\mathbf{H}' = \mathbf{P'\Theta'\Psi}$ as (21), there are $C_{ZN}^{s'}$ submatrices $\mathbf{H}_I'$ obtained by randomly selecting columns of size $s'$. The Gram matrix $G(\mathbf{H}')$ is given by

$$G(\mathbf{H}') = G(\mathbf{H}_1) + G(\mathbf{H}_2) + \cdots + G(\mathbf{H}_P) \quad (29)$$

which contains all the eigenvalue information of its submatrices. The eigenvalues of each submatrix distributed in $P[1-\delta_{s'}, 1+\delta_{s'}]$ are the same complicated permutation and combination problem as the Gram matrix of $\mathbf{H}$. Meanwhile, the eigenvalues of each submatrix distributed in $[1-\delta_{s'}, 1+\delta_{s'}]$ of the normalized Gram matrix $G'(\mathbf{H}') = G(\mathbf{H}')/P$ is used to be discussed in the following for facilitative comparative analysis based on the Gershgorin circle theorem.

Accordingly, the diagonal elements of the Gram matrix $G(\mathbf{H})$ and the normalized Gram matrix $G'(\mathbf{H}')$ are equal to one identically, which means that the sensing matrix $\mathbf{H}$ and $\mathbf{H}'$ meet the first-order RIP and the received signal can be recovered with high fidelity when the input is first-order sparsity.

Then, for the second-order sparsity case, by choosing $\delta_0, \delta_d > 0$ appropriately and $\delta_0 + \delta_d = \delta_s \in (0,1)$, the diagonal elements $G_{i,i}$ of the Gram matrix $G(\mathbf{H})$ and the off-diagonal elements satisfy $|G_{i,i} - 1| = 0 < \delta_0$ and $|G_{i,j} - 1| < \delta_d/s$, which is the same for the normalized Gram matrix $G'(\mathbf{H}')$. It means that the center of the Gershgorin circle for the Gram matrix deviates from the origin of the coordinate axis by $1 \pm \delta_0$ and the radius is less than $(s-1)\delta_d/s < \delta_d$. The eigenvalues of the Gram matrix are in the range $(1 - \delta_0 - \delta_d, 1 + \delta_0 + \delta_d) = (1 - \delta_s, 1 + \delta_s)$ and the sensing matrix meets the second-order RIP. Therefore, there is just a matter of consideration to prove that each of the non-diagonal elements for the Gram matrix is less than one.

The elements of the matrix $\mathbf{T}_{ij}$ can be written as

$$a_{r,c} = \frac{1}{N} \sum_{n=1}^{N} e^{-jM_{ij}A_\theta\theta(nT)} \psi^{(n-1)(r-c)} \quad (30)$$

and its module can be expressed as

$$|a_{r,c}| = \left|\frac{1}{N} \sum_{n=1}^{N} e^{-jM_{ij}A_\theta\theta(nT)} \psi^{(n-1)(r-c)}\right|$$
$$\leq \frac{1}{N} \begin{pmatrix} \left|\sum_{n=1}^{N} e^{-jM_{ij}A_\theta\theta(nT)}\right| + \\ \left|\sum_{n=1}^{N} e^{-jM_{ij}A_\theta\theta(nT)} \psi^{(r-c)}\right| + \\ \cdots + \left|\sum_{n=1}^{N} e^{-jM_{ij}A_\theta\theta(nT)} \psi^{(N-1)(r-c)}\right| \end{pmatrix} = 1 \quad (31)$$

where $|a_{r,c}| = 1$ if and only if $\theta(nT) = 0$ for $n = 1, 2, \ldots, N$. Thus $|a_{r,c}| < 1$, meaning that the eigenvalues of the Gram matrix composed of any two columns of the sensing matrix are in the range $(0,1)$, which is the same for the normalized Gram matrix $G'(\mathbf{H}')$.

Therefore, the sensing matrix $\mathbf{H}$ and $\mathbf{H}'$ meet the second-order RIP. Further analysis related to system parameter settings for the higher-order RIP case is discussed in section IV.

$$\mathbf{T}_{ij} = \frac{1}{N} \begin{bmatrix} \sum_{n=1}^{N} e^{-jM_{ij}A_{\theta}\theta(nT)} & \sum_{n=1}^{N} e^{-jM_{ij}A_{\theta}\theta(nT)} \psi^{(n-1)} & \cdots & \sum_{n=1}^{N} e^{-jM_{ij}A_{\theta}\theta(nT)} \psi^{(n-1)(N-1)} \\ \sum_{n=1}^{N} e^{-jM_{ij}A_{\theta}\theta(nT)} \psi^{-(n-1)} & \sum_{n=1}^{N} e^{-jM_{ij}A_{\theta}\theta(nT)} & \cdots & \sum_{n=1}^{N} e^{-jM_{ij}A_{\theta}\theta(nT)} \psi^{(n-1)(N-2)} \\ \vdots & \vdots & \ddots & \vdots \\ \sum_{n=1}^{N-1} e^{-jM_{ij}A_{\theta}\theta(nT)} \psi^{-(n-1)(N-1)} & \sum_{n=1}^{N} e^{-jM_{ij}A_{\theta}\theta(nT)} \psi^{-(n-1)(N-2)} & \cdots & \sum_{n=1}^{N} e^{-jM_{ij}A_{\theta}\theta(nT)} \end{bmatrix} \quad (32)$$

*C. The BRIP analysis*

The wide bandwidth signals can be modeled as block sparse signals, which are more suitable for BRIP analysis. Considering the vector $\mathbf{X}$ can be divided into $L$ blocks on the length index set of the subblock

$$D = \{d_1, d_2, \ldots, d_L\} \quad (33)$$

where the length of each block can be different, and there can be overlapping parts between the subblocks. Denoting that $\mathbf{X}[l]$ is the $l$ th sub-block of length $d_l$, and then the $\mathbf{X}$ is

$$\mathbf{X}^{\mathrm{T}} = [\underbrace{X_1 \ldots X_{d_1}}_{\mathbf{x}[1]} \ldots \underbrace{X_{d_{l-1}+1} \ldots X_{d_l}}_{\mathbf{x}[l]} \ldots \underbrace{X_{ZN-d_L+1} \ldots X_{ZN}}_{\mathbf{x}[L]}] \quad (34)$$

where $\mathbf{X}$ is the DFT vector of the Nyquist-rate sampled for the conventional sparse signal when $d_l = 1$. Similarly, the sensing matrix $\mathbf{H}$ can be represented as a concatenation of the column block matrix $\mathbf{H}[l]$ of size $N \times d_l$, as follows

$$\mathbf{H} = [\underbrace{\mathbf{h}_1 \ldots \mathbf{h}_{d_1}}_{\mathbf{H}[1]} \ldots \underbrace{\mathbf{h}_{d_{l-1}+1} \ldots \mathbf{h}_{d_l}}_{\mathbf{H}[l]} \ldots \underbrace{\mathbf{h}_{ZN-d_L+1} \ldots \mathbf{h}_{ZN}}_{\mathbf{H}[L]}] \quad (35)$$

where $\mathbf{h}_i$ denotes the column vector of the sensing matrix $\mathbf{H}$ for $i = 1, 2, \ldots, ZN$.

The BRIP is then defined as follows: the sensing matrix $\mathbf{H}$ is said to satisfy the BRIP with parameters $(s_B, \delta_B)$ for $s_B \leq L, 0 \leq \delta_B \leq 1$ over the subblock index set $D$, if for all subblock index set $I_B \subset \{1, 2, \ldots, L\}$ of $\mathbf{H}$ such that $|I_B| \leq s_B$, and for all $\boldsymbol{\theta}_B \in \mathbb{C}^{\sum_{l \in I_B} d_l}$, one has

$$(1-\delta_B)\|\boldsymbol{\theta}_B\|_2^2 \leq \|\mathbf{H}_{I_B} \boldsymbol{\theta}_B\|_2^2 \leq (1+\delta_B)\|\boldsymbol{\theta}_B\|_2^2 \quad (36)$$

where $\mathbf{H}_{I_B}$ is the submatrix of $\mathbf{H}$ consisting of the related subblock specified by the subblock index set $I_B \subset \{1, 2, \ldots, L\}$ of size $N \times \sum_{l \in I_B} d_l$. Then, the parameter $s_B$ denotes the block sparsity and the minimum of all $\delta_B$ is the block restricted isometry constant (BRIC) $\delta_{s_B|D}$. There is an inequality relation between the BRIC and the eigenvalues of the matrix $\mathbf{H}_{I_B}^{\mathrm{H}} \mathbf{H}_{I_B}$, as follows

$$(1-\delta_{s_B})\|\boldsymbol{\theta}_B\|_2^2 \leq \lambda_{\min}(\mathbf{H}_{I_B}^{\mathrm{H}} \mathbf{H}_{I_B})\|\boldsymbol{\theta}_B\|_2^2 \leq \|\mathbf{H}_{I_B} \boldsymbol{\theta}_B\|_2^2 \\ \leq \lambda_{\max}(\mathbf{H}_{I_B}^{\mathrm{H}} \mathbf{H}_{I_B})\|\boldsymbol{\theta}_B\|_2^2 \leq (1+\delta_{s_B})\|\boldsymbol{\theta}_B\|_2^2 \quad (37)$$

where $\lambda_{\min}(\mathbf{H}_{I_B}^{\mathrm{H}} \mathbf{H}_{I_B})$ and $\lambda_{\max}(\mathbf{H}_{I_B}^{\mathrm{H}} \mathbf{H}_{I_B})$ denote the minimal and maximal eigenvalues of $\mathbf{H}_{I_B}^{\mathrm{H}} \mathbf{H}_{I_B}$, respectively.

Then the Gershgorin circle theorem is also used to evaluate the BRIP. Through the block coherence constant $\upsilon_{I_B}$ of the subblock matrix on the diagonal of the Gram matrix and the block coherence constant $\mu_{I_B}$ of the subblock matrix on the non-diagonal of the Gram matrix, the inequality (37) can be changed into

$$\|\mathbf{H}_{I_B} \boldsymbol{\theta}_B\|_2^2 = \boldsymbol{\theta}_B^{\mathrm{H}} \mathbf{H}_{I_B}^{\mathrm{H}} \mathbf{H}_{I_B} \boldsymbol{\theta}_B = \sum_{c=1}^{s_B} \sum_{r=1}^{s_B} \boldsymbol{\theta}_B^{\mathrm{H}}[c] \mathbf{M}[c,r] \boldsymbol{\theta}_B[r] \\ = \sum_{c=1}^{s_B} \boldsymbol{\theta}_B^{\mathrm{H}}[c] \mathbf{M}[c,c] \boldsymbol{\theta}_B[r] \\ + \sum_{c=1}^{s_B} \sum_{r=1, r \neq c}^{s_B} \boldsymbol{\theta}_B^{\mathrm{H}}[c] \mathbf{M}[c,r] \boldsymbol{\theta}_B[r] \\ \leq \|\boldsymbol{\theta}_B\|_2^2 + \sum_{c=1}^{s_B} \max \rho(\mathbf{M}[c,c] - \mathbf{I}_d) \|\boldsymbol{\theta}_B\|_2^2 \\ + \sum_{c=1}^{s_B} \sum_{r=1, r \neq c}^{s_B} \max \rho(\mathbf{M}[c,r]) \boldsymbol{\theta}_B^{\mathrm{H}}[c] \boldsymbol{\theta}_B[r] \\ \leq [1 + (d-1)\upsilon_{I_B} + (I_B - 1)d\mu_{I_B}] \|\boldsymbol{\theta}_B\|_2^2 \quad (38)$$

and

$$\|\mathbf{H}_{I_B} \boldsymbol{\theta}_B\|_2^2 = \boldsymbol{\theta}_B^{\mathrm{H}} \mathbf{H}_{I_B}^{\mathrm{H}} \mathbf{H}_{I_B} \boldsymbol{\theta}_B = \sum_{c=1}^{s_B} \sum_{r=1}^{s_B} \boldsymbol{\theta}_B^{\mathrm{H}}[c] \mathbf{M}[c,r] \boldsymbol{\theta}_B[r] \\ = \sum_{c=1}^{s_B} \boldsymbol{\theta}_B^{\mathrm{H}}[c] \mathbf{M}[c,c] \boldsymbol{\theta}_B[r] \\ + \sum_{c=1}^{s_B} \sum_{r=1, r \neq c}^{s_B} \boldsymbol{\theta}_B^{\mathrm{H}}[c] \mathbf{M}[c,r] \boldsymbol{\theta}_B[r] \\ \geq \|\boldsymbol{\theta}_B\|_2^2 - \sum_{c=1}^{s_B} \max \rho(\mathbf{M}[c,c] - \mathbf{I}_d) \|\boldsymbol{\theta}_B\|_2^2 \\ - \sum_{c=1}^{s_B} \sum_{r=1, r \neq c}^{s_B} \max \rho(\mathbf{M}[c,r]) \boldsymbol{\theta}_B^{\mathrm{H}}[c] \boldsymbol{\theta}_B[r] \\ \geq [1 - (d-1)\upsilon_{I_B} - (I_B - 1)d\mu_{I_B}] \|\boldsymbol{\theta}_B\|_2^2 \quad (39)$$

where $\mathbf{M}[c,r] = \mathbf{H}_{I_B}^{\mathrm{H}}(c) \mathbf{H}_{I_B}(r)$.

Thus, the BRIC can be expressed as follows

$$\delta_{s_B|D} = (d-1)\upsilon_{I_B} + (I_B - 1)d\mu_{I_B} \quad (40)$$

Furthermore, for the UAV swarm using NYFR, the number of elements for each Nyquist zone is the same. Therefore, by selecting each Nyquist zone as a block with a fixed length of $N$ points, then there are $Z$ blocks. Moreover, the sensing matrix is also divided into $Z$ blocks, where each block contains $N$ column vectors. As a result, the sensing matrix $\mathbf{H}$ and $\mathbf{H}'$ meet the first-order BRIP based on that the minimal block coherence constant $\upsilon_{I_B}$ of the subblock matrix on the diagonal. For the second-order block sparsity case, because of

$$\mathbf{M}[c,r] = \mathbf{T}_{(c-1)(r-1)} \qquad (41)$$

where $c, r = 1, 2, \ldots, Z$ and $c \neq r$. Since the $\mathbf{T}_{(c-1)(r-1)}$ is a Toeplitz matrix from (30), which satisfies the RIP, namely that there is $\lambda_{\max}^{1/2} = \left(\mathbf{T}_{(c-1)(r-1)}^{H} \mathbf{T}_{(c-1)(r-1)}\right) \in [0,1]$. Thus, the sensing matrix $\mathbf{H}$ and $\mathbf{H}'$ meet the second-order BRIP as $d\mu_{I_B} \in (0,1)$, and the block CS techniques can be used to process the wide bandwidth signals. Further analysis related to system parameter settings for the higher-order BRIP case is discussed in section IV.

## IV. SIMULATION AND DISCUSSION

In this section, the availability of the sensing model and the analysis of RIP and BRIP are discussed through simulation for the two scenarios described in part II. Meanwhile, the wideband spectrum recovery for the distributed UAV swarm is displayed by the following.

### A. Sensing Model Analysis for the UAV Swarm with RIS

With the simulation parameters in Table I, there are 16 UAVs with RIS forming a group, and the distance between adjacent UAVs satisfies the half-wavelength requirement. Additionally, only 0 or $\pi$ phases can be controlled for the RIS in the measurement matrix. Meanwhile, the NYFR system covers 4 Nyquist zones with a bandwidth of 4GHz for each one. There are 800 measurements and 3200 Nyquist samples. Then the amplitude and frequency of the sinusoidal modulation are one and 5MHz, respectively.

First, the distribution diagram of elements for the Gram matrix $G(\mathbf{H})$ is shown in Fig. 5. The main diagonal elements are one, and the non-diagonal elements are limited in $(0,1)$, which has the same conclusion in Part III-B. The system satisfies the first-order and second-order RIP based on the parameters given in Table I. However, the sensing matrix does not meet the third-order RIP or more. It is worth noting that the non-zero elements of the non-diagonal are located at the minor diagonal by the intervals of $N$.

TABLE I
SIMULATION PARAMETERS

| Parameter | Symbol | Value |
|---|---|---|
| The number of UAV swarm with RIS | $M$ | 16 |
| The distance between adjacent UAVs | $d$ | $\lambda_{\max}/2$ |
| The sampling rate of ADC | $f_s$ | 4GHz |
| The number of Nyquist zone | $Z$ | 4 |
| The frequency of the phase modulation signal | $f_\theta$ | 5MHz |
| The amplitude of the phase modulation signal | $A_\theta$ | 1 |
| The number of measurements | $N$ | 800 |

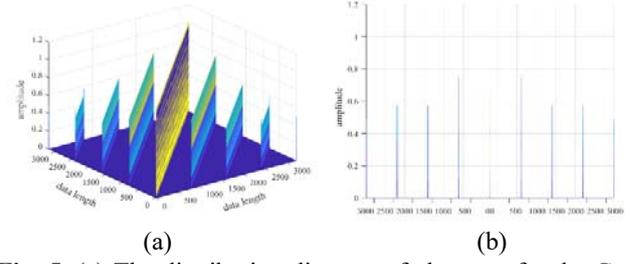

**Fig. 5.** (a) The distribution diagram of elements for the Gram matrix $G(\mathbf{H})$. (b) The sectional view at azimuth -45°.

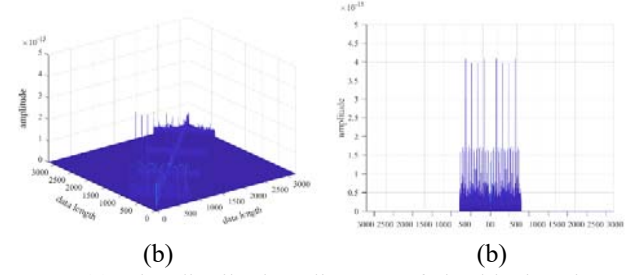

**Fig. 6.** (a) The distribution diagram of the block coherence constant $\upsilon_{I_B}$ for the subblock matrix on the diagonal of the Gram matrix. (b) The sectional view at azimuth -45°.

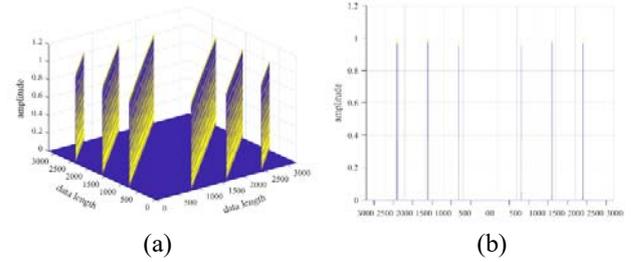

**Fig. 7.** (a) The distribution diagram of the block coherence constant $N\mu_{I_B}$ for the subblock matrix on the non-diagonal of the Gram matrix $\mathbf{T}_{(c-1)(r-1)}^{H}\mathbf{T}_{(c-1)(r-1)}$. (b) The sectional view at azimuth -45°.

It means that the frequencies of interest around multiples of the folded sampling rate $f_s$ have a low recovery probability, which directly limits the RIP for the system. Meanwhile, for the other frequencies except that, the system can meet a higher-order RIP. Therefore, the BRIP analysis is more applicable to this issue. Thus, by partitioning the Nyquist zone into units, the elements of the block coherence constant $\upsilon_{I_B}$ for the subblock matrix on the diagonal of the Gram matrix are approximately zero, as shown in Fig. 6, which means that the system meets the first-order BRIP.

Then Fig. 7 shows the distribution diagram of the block coherence constant $N\mu_{I_B}$ for the subblock matrix on the non-diagonal of the Gram matrix $\mathbf{T}_{(c-1)(r-1)}^{H}\mathbf{T}_{(c-1)(r-1)}$. We can see that the diagonal elements are one, and the other non-diagonal elements within each subblock are about zero. Thus, there is $\delta_{s_B|D} \in (0,1)$ for the second-order block sparsity, but the system cannot meet the higher-order BRIP.

Furthermore, Fig. 8 displays the correlation of the sensing matrix under the different system parameters by using the maximum value of the non-diagonal elements for the Gram matrix. From this, it is apparent that a sensing matrix with a larger amplitude $A_\theta$ of LO can be obtained, resulting in smaller correlation and better reconstruction performance. However, the correlation does not blindly increase with the frequency $f_\theta$ for the phase modulation signal of LO. In this figure, there is little difference in correlation change for the frequencies of 5M to 30M. Thus, there is a trade-off between resources and the system quality.

*B. Sensing Model Analysis for the Decentralized UAV Swarm*

In the simulation discussed above, a scenario with three UAVs is considered, and the distance between the adjacent UAVs is within 15km. The NYFR system for each UAV covers 4 Nyquist zones with a bandwidth of 4GHz for each one. There are 800 measurements and 3200 Nyquist samples, which is the same as listed in Table I.

Meanwhile, the amplitude and frequency of the modulation for each UAV are different. First, there is the distribution diagram of elements for the Gram matrix with amplitude and frequency as $A_{\theta_1}=1$, $f_{\theta_1}=5\text{MHz}$, $A_{\theta_2}=1, f_{\theta_2}=10\text{MHz}$ and $A_{\theta_3}=1, f_{\theta_3}=30\text{MHz}$, as shown in Fig. 9.

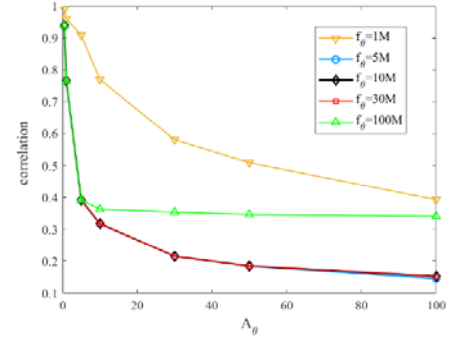

**Fig. 8.** The correlation comparison for sensing matrix under different system parameters.

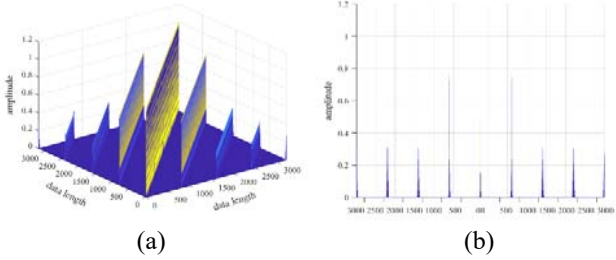

**Fig. 9.** (a) The distribution diagram of elements for the Gram matrix with $A_{\theta_1}=1, f_{\theta_1}=5\text{MHz}$, $A_{\theta_2}=1, f_{\theta_2}=10\text{MHz}$ and $A_{\theta_3}=1, f_{\theta_3}=30\text{MHz}$. (b) The sectional view at azimuth -45°.

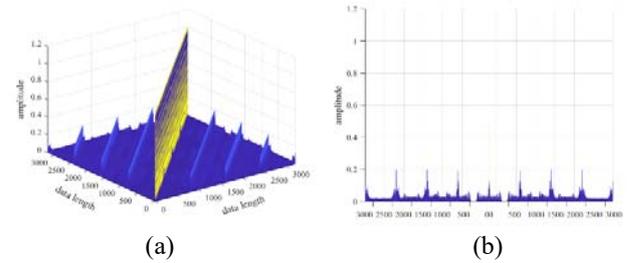

**Fig. 12.** (a) The distribution diagram of elements for the Gram matrix with $A_{\theta_1}=1, f_{\theta_1}=5\text{MHz}$, $A_{\theta_2}=10, f_{\theta_2}=10\text{MHz}$ and $A_{\theta_3}=50, f_{\theta_3}=30\text{MHz}$. (b) The sectional view at azimuth -45°.

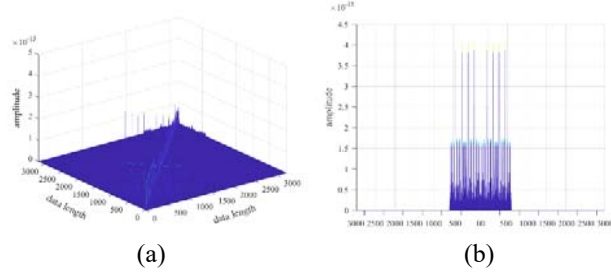

**Fig. 10.** (a) The distribution of the block coherence constant $\upsilon_{I_B}$ with $A_{\theta_1}=1, f_{\theta_1}=5\text{MHz}$, $A_{\theta_2}=1, f_{\theta_2}=10\text{MHz}$ and $A_{\theta_3}=1, f_{\theta_3}=30\text{MHz}$. (b) The sectional view at azimuth -45°.

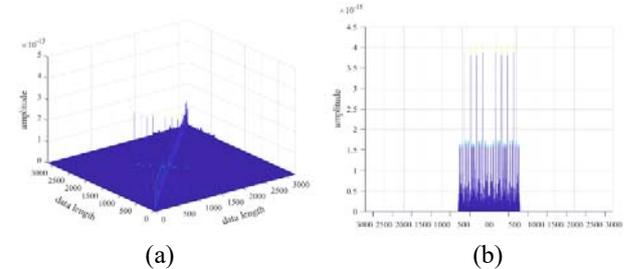

**Fig. 13.** (a) The distribution of the block coherence constant $\upsilon_{I_B}$ with $A_{\theta_1}=1, f_{\theta_1}=5\text{MHz}$, $A_{\theta_2}=10, f_{\theta_2}=10\text{MHz}$ and $A_{\theta_3}=50, f_{\theta_3}=30\text{MHz}$. (b) The sectional view at azimuth -45°.

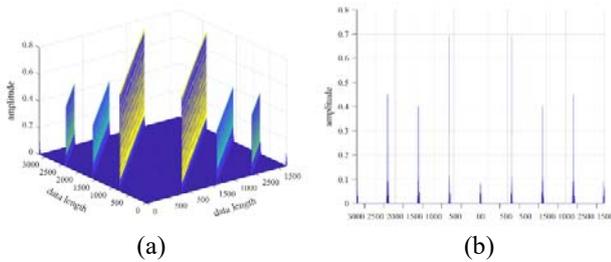

**Fig. 11.** (a) The distribution of the block coherence constant $N\mu_{I_B}$ with $A_{\theta_1}=1, f_{\theta_1}=5\text{MHz}$, $A_{\theta_2}=1, f_{\theta_2}=10\text{MHz}$ and $A_{\theta_3}=1, f_{\theta_3}=30\text{MHz}$. (b) The sectional view at azimuth -45°.

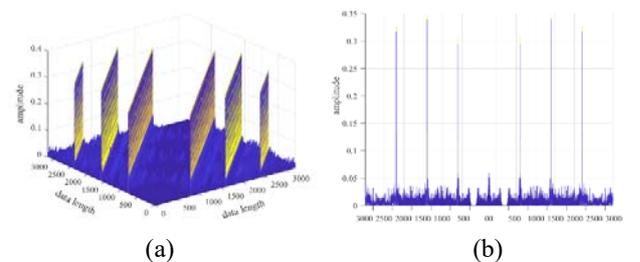

**Fig. 14.** (a) The distribution of the block coherence constant $N\mu_{I_B}$ with $A_{\theta_1}=1, f_{\theta_1}=5\text{MHz}$, $A_{\theta_2}=10, f_{\theta_2}=10\text{MHz}$ and $A_{\theta_3}=50, f_{\theta_3}=30\text{MHz}$. (b) The sectional view at azimuth -45°.

The correlation of the sensing matrix is the same for each single UAV system as discussed in Fig. 8. However, frequencies of interest around multiples of the folded sampling rate have a better recovery probability than in Fig. 5, based on more incoherent samples for the sensing model of the decentralized UAV swarm. Moreover, the distribution of the block coherence constant is the same as in Fig. 6 when taking the Nyquist zone as a unit, as shown in Fig. 10. While, there is a better quality for the block coherence constant in Fig. 11 compared to Fig. 7. Therefore, according to the analysis of the sensing model for the decentralized UAV swarm, incoherent measurements from the UAV swarm can reconstruct signals that are less sparse.

Furthermore, the simulation results of RIP and BRIP are shown in Fig. 12 to Fig. 14 for the decentralized UAV swarm with different modulation amplitude compared to Fig. 9 to Fig. 11. Notably, the non-diagonal elements of the Gram matrix are limited to (0, 0.2) in Fig. 12, which indicates that the system meets the fifth-order RIP. And for the other frequencies except that around multiples of the folded sampling rate, the system can meet a higher-order RIP towards random sampling, as shown in Fig. 15. Then, the performance of block sparse signal reconstruction is improved as Fig. 14. Thus, as the incoherent samples increase, the ability to reconstruct signals of less sparse is improved.

*C. Spectrum Recovery for the Distributed UAV Swarm*

Under the same simulation parameters listed in Table I, there are 3 UAVs assumed to form a distributed UAV swarm with the distance between the adjacent UAVs satisfying the half-wavelength. Additionally, the amplitude and frequency for the phase modulation signal of the swarm are [1, 10MHz], [30, 10MHz] and [30, 30MHz], respectively. Afterward, the Pearson correlation coefficient (PCC) between the recovered signals and raw signals is adopted to evaluate the performance, defined as

$$\mathrm{PCC}(\hat{\mathbf{X}},\mathbf{X}) = \frac{\sum_{i=1}^{ZN}(\hat{\mathbf{X}}_i - \mu_{\hat{\mathbf{X}}})(\mathbf{X}_i - \mu_{\mathbf{X}})}{\sqrt{\sum_{i=1}^{ZN}(\hat{\mathbf{X}}_i - \mu_{\hat{\mathbf{X}}})^2}\sqrt{\sum_{i=1}^{ZN}(\mathbf{X}_i - \mu_{\mathbf{X}})^2}} \quad (42)$$

where $\mu_{\hat{\mathbf{X}}}$ and $\mu_{\mathbf{X}}$ denote the mean value of the recovered signals $\hat{\mathbf{X}}$ and raw signals $\mathbf{X}$, respectively.

The reconstruction performance for pulses is of particular interest in radar and communication systems. The pulse length requirement of received signals is considered first for reconstruction. As shown in Fig. 16, there is $K=1$ frequency component which is selected in the 2-18GHz randomly, with an input SNR set to 0dB. Moreover, the start time of the pulses is randomly distributed throughout the observation duration. The PCC results under a UAV swarm with RIS based on different phase modulation signals and a distributed UAV swarm based on their combination are compared. The PCC increases with pluses length based on one hundred Monte Carlo trials. The average PCC reaches more than 90% when the pulse length is 150ns.

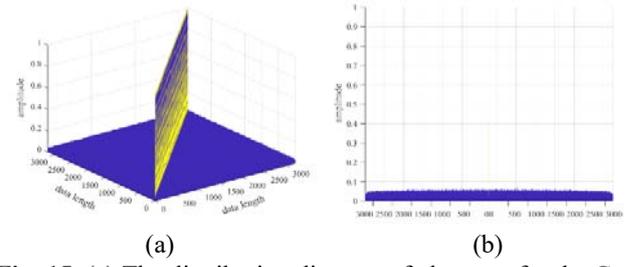

**Fig. 15.** (a) The distribution diagram of elements for the Gram matrix with random. (b) The sectional view at azimuth -45°.

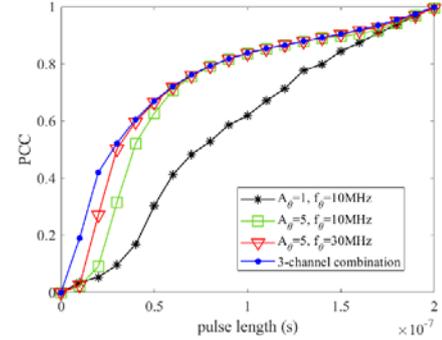

**Fig. 16.** Pulse length requirement for reconstruction ( $K = 1$, SNR = 0dB).

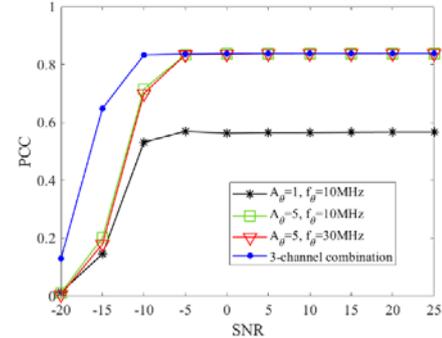

**Fig. 17.** Reconstruction performance versus SNR ( $K = 1$ ).

Meanwhile, as $A_\theta$ increases, PCC gets bigger, while as $f_\theta$ increases, PCC changes little. This is consistent with the conclusion from Fig. 8. However, the combination of three channels provides more gain. That is because there is consideration of spatial gain, but the multi-channel combination has little advantage in processing the single signal.

In Fig. 17, PCC results are compared as a function of the input SNR, where $K=1$ is assumed and randomly selected in the 2-18GHz. Meanwhile, the pulse length is fixed at 100ns with a start time randomly distributed throughout the observation duration. It can be seen that all results tend to be stabilized when SNR is greater than -5dB. And the combination of three channels obtains an extra 5dB.

Similarly, PCC results are compared as a function of sparsity with an SNR of 0dB and pulse length of 200ns. As seen in Fig. 18, the lower correlation of the sensing matrix for a UAV swarm, the less reconstruction performance is affected by the sparsity. Here the advantage of a distributed UAV swarm is clearly reflected. Moreover, there are bound to be different Nyquist zones of received signals folded into the same baseband frequency, for wideband spectrum sensing based on sub-Nyquist sampling. Similarly, there are also cases where the mirror frequencies fold into the same baseband frequency.

Therefore, the reconstruction performance versus the same aliasing is shown in Fig. 19. In this simulation, the frequency components consist of folding units based on the sparsity, randomly selected with the addition or subtraction of only one frequency in the 2-4GHz. And the folding units are integral multiples of $f_s$. For example, a frequency is selected 2.1GHz which is folded into 0.9GHz as a mirror component of the baseband. Then the same aliasing frequencies have 6.1GHz, 10.1GHz, and 14.1GHz as mirror components and others such as 5.9GHz, 9.9GHz, 13.9GHz, and 15.9GHz. Moreover, there are different input frequencies completely aliased into the same baseband frequency at most twice the number of Nyquist zones. Therefore, the sparsity increases in Fig. 18 when only randomly selected in the same aliasing frequencies. Thus, the PCC results tend to be stable.

Furthermore, the multiple signals reconstruction spectrum is displayed in Fig. 20 with the 0dB SNR. There are ten inputs with frequency components randomly selected in the 2-18GHz, including a continuous signal, two binary phase shift keying (BPSK) signals with random code, two linear frequency modulation (LFM) signals with a bandwidth of 20MHz, and five monopulse signals whose pulse length and start time are randomly distributed throughout the observation duration. From the spectrogram of the NYFR output, there is a sinusoidal frequency modulation for each received signal, and the degree of the modulation depends on the Nyquist zone where the signals are located. In addition, a distributed UAV swarm augments wideband spectrum acquisition capabilities compared to the others.

On the other hand, for wide bandwidth signals, the reconstruction performance versus bandwidth of the received signal is discussed based on block recovery methods in Fig. 21, which is compared to non-block recovery methods. There is only one LFM signal considered with a pulse length of 200ns, whose carrier frequency is randomly selected in the 2-18GHz, where the input SNR is set to 0dB.

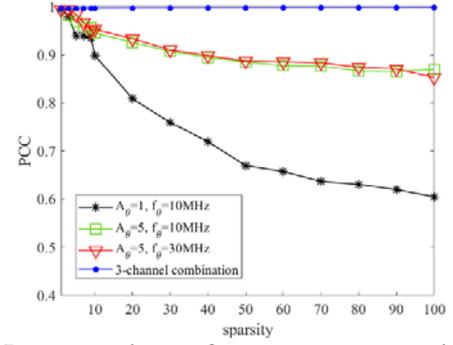

**Fig. 18.** Reconstruction performance versus sparsity ( SNR = 0dB ).

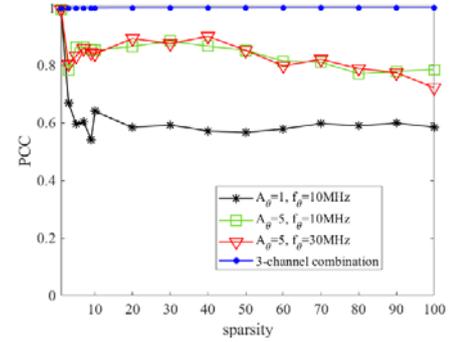

**Fig. 19.** Reconstruction performance versus same aliasing ( SNR = 0dB ).

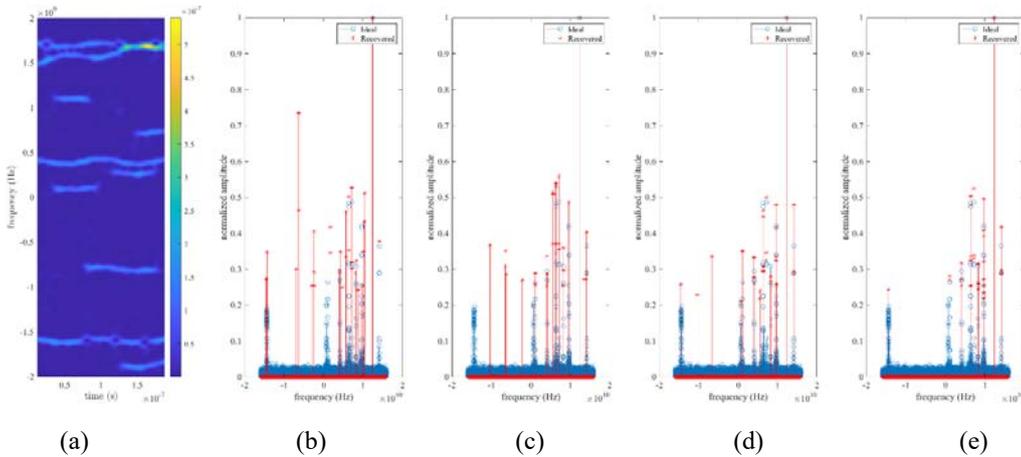

(a)          (b)          (c)          (d)          (e)

**Fig. 20.** Multiple signals reconstruction spectrum ( SNR = 0dB ). (a) Spectrogram of the NYFR output. (b) Spectrum of the UAV swarm with RIS based $A_\theta = 1, f_\theta = 10\text{MHz}$. (c) Spectrum of the UAV swarm with RIS based $A_\theta = 5, f_\theta = 10\text{MHz}$. (d) Spectrum of the UAV swarm with RIS based $A_\theta = 5, f_\theta = 30\text{MHz}$. (e) Spectrum of the distributed UAV swarm based on the combination of (b)~(d).

As we can see that the harmonic components introduced are much more numerous than those with larger sparsity for inputs with very wide bandwidth, and the performance of the non-block recovery algorithm decreases rapidly compared to Fig. 18. That is because that not only introduces more related folding noise but also brings ambiguity problem for cross-NZ signals. And by modeling the wide bandwidth signals as block sparse input, the recovery performance is consistent with the analysis in the section III-C.

In the end, the influence of the phase matching degree between the sensing model and measurements on the reconstruction performance is discussed in Fig. 22 with the 0dB SNR. There is only one input received for different phase drift, whose carrier frequency is randomly selected in the 2-18GHz with a pulse length of 200ns. The recovery procedure simulates time requirements for the unknown alignment of LO and ADC based on the constant phase of the sensing matrix. As a result, the model matching degree is more influenced by magnitude. And the maximum drift within them is the same for multi-channel combination cases, where the delay must be within 7ns for the phase modulation of $A_\theta = 1, f_\theta = 10\text{MHz}$.

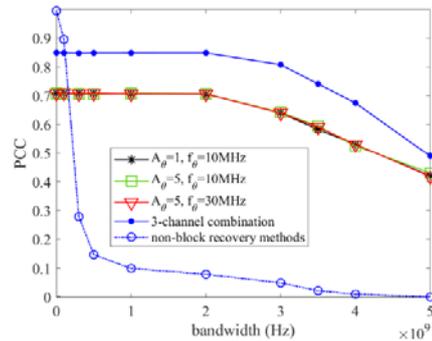

**Fig. 21.** Reconstruction performance versus bandwidth ( K = 1, SNR = 0dB ).

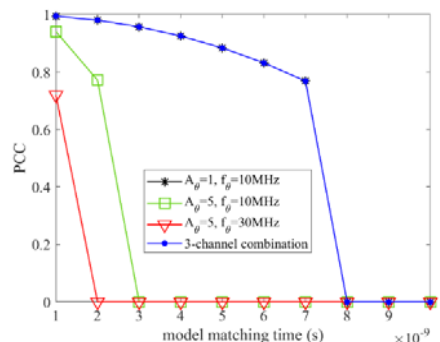

**Fig. 22.** Reconstruction performance versus model matching ( K = 1, SNR = 0dB ).

## V. Conclusion

Distributed UAV swarm is composed of multiple flexible and intelligent UAVs, which can collaboratively perform missions more reliably, economically, and efficiently. While the potentially dense adoption of UAV swarms and the bandwidth-hungry nature of their applications will undoubtedly increase the shortage of spectrum resources. In this paper, two multichannel scenarios for the distributed UAV swarms are considered to augment wideband spectrum sensing using the NYFR architecture, one with a complete functional receiver for the UAV swarm with RIS, and another with a decentralized UAV swarm equipped with a complete functional receiver for each UAV element. Different from the others, the multichannel sensing model of CS for distributed UAV swarms in a wideband spectrum is formulated. Further, the RIP of the multichannel sensing model for multiple pulse reconstruction is analyzed, particularly for short pulses which can be recovered as the boundary conditions. Additionally, the BRIP is analyzed for the wide bandwidth signals and cross-NZ signals based on the block sparse recovery. The proposed technology can improve processing capability for multiple signals and cross-NZ signals while reducing interference from folded noise and harmonics. Experiment results demonstrate augmented spectrum sensing efficiency under the non-strictly sparse condition.

## Acknowledgment

The authors would like to thank the reviewers for their constructive comments.